\documentclass{article}
\usepackage{ijcai19}

\usepackage{times}
\usepackage{soul}
\usepackage{url}
\usepackage[hidelinks]{hyperref}
\usepackage[utf8]{inputenc}
\usepackage[small]{caption}
\usepackage{graphicx}
\usepackage{amsmath}
\usepackage{booktabs}
\usepackage{cite}
\usepackage{algorithm}
\usepackage{algpseudocode}
\usepackage{multirow}
\title{Learning Sound Events from Webly Labeled Data}

  \author{
  Anurag Kumar\and
  Ankit Shah\and
  Alexander Hauptmann \And Bhiksha Raj\\
  \affiliations
    Language Technologies Institute, School of Computer Science, Carnegie Mellon University\\
  \emails
 argxkr@gmail.com,
  \{aps1, alex, bhiksha\}@cs.cmu.edu
 }

\begin{document}

\maketitle

\begin{abstract}
In the last couple of years, weakly labeled learning has turned out to be an exciting approach for audio event detection. In this work, we introduce \emph{webly labeled} learning for sound events which aims to remove human supervision altogether from the learning process. We first develop a method of obtaining labeled audio data from the web (\emph{albeit noisy}), in which no manual labeling is involved. We then describe methods to efficiently learn from these webly labeled audio recordings. In our proposed system, \emph{WeblyNet}, two deep neural networks \emph{co-teach} each other to robustly learn from webly labeled data, leading to around \textbf{17\%} relative improvement over the baseline method. The method also involves transfer learning to obtain efficient representations.
\end{abstract}

\section{Introduction}
\label{sec:intro}

As artificial intelligence becomes an increasingly integral part of our life, it is imperative that automated understanding of sounds too gets integrated into everyday systems and devices. Sound event detection and understanding has a wide range of applications \cite{kumar2018thesis}, and hence, in the past few years, the field has received considerable attention in the broader areas of machine learning and audio processing.

One long-standing problem in audio event detection (AED) has been the availability of labeled data. Labeling sound events in an audio stream require marking their beginnings and ends. Annotating audio recordings with times of occurrences of events is a laborious and resource intensive task. Weakly-supervised learning for sound events \cite{kumar2016audio} addressed this issue by showing that it is possible to train audio event detectors using \emph{weakly labeled} data: audio recordings, here, are tagged only with presence or absence of the events as opposed to the time stamp annotations in  \emph{strongly labeled} audio data. 

Weakly labeled AED has gained significant attention since it was first proposed and has become the preferred and the most promising approach for scaling AED. Several weakly labeled methods have been proposed in last couple of years {\em e.g.} \cite{kumar2017knowledge, chou2018learning,  mcfee2018adaptive,lu2018temporal, xu2017attention}, to mention a few. Weak labeling has enabled the collection of audio-event datasets much larger than before  \cite{gemmeke2017audio,fonseca2017freesound}. Moreover, learning from weak labels features in the annual DCASE challenge for sound events detection as well \footnote{http://dcase.community/}.

Being able to work with weak labels is, however, only half the story.  Even weak labeling, when done manually, becomes challenging on large scale; tagging a large number of audio recordings for a large number of sound classes is non-trivial. Datasets along the lines of AudioSet \cite{gemmeke2017audio} are not easy to create and require considerable resources.  However, a big advantage of weakly labeled learning is that it opens up the possibility of learning from the data on the web \emph{without employing manual annotation}, thereby allowing large scale learning without laborious human supervision. 
%

The web provides us with a rich resource from which weakly-labeled data could be easily derived. It removes the resource intensive processing of creating the training data manually and opens up the possibility of completely automated training. However, this brings up a new problem -- the weak labels associated with these recordings, having been automatically obtained through some means, are likely to be noisy. The challenge now extends to being able to learn from {\em weakly} and {\em noisily} labeled web data.  We call such data \textbf{\emph{webly labeled}}. This paper proposes solutions to learning from webly-labeled data.

There have been several works on webly supervised learning of visual objects and concepts \cite{chen2015webly, liang2016learning, divvala2014learning}. However, learning sound events from webly labeled data has received little to no attention. The main prior work here is \cite{kumar2017audio}, where webly labeled data have been employed; however, to counter the noise in the labels, strongly labeled data is used to provide additional supervision. Needless to say, the strong labels which act as the supporting data are manually obtained.

Our objective in this paper is to eliminate human supervision altogether from the learning process by proposing webly supervised learning of sounds. Webly labeled data by default are weakly labeled and hence our proposed methods are designed for weakly labeled audio recordings. Our motivation then is to introduce a learning scheme which can effectively counter  additional challenges of webly labeled data. We first present the challenges of webly labeled learning of sounds and then an outline of the proposed system  in next section.  


\subsection{Challenges in Webly Labeled Learning}
Webly labeled learning involves several challenges. The first one is obtaining the webly labeled data itself. The challenge of obtaining quality exemplars from the web has been well documented in several computer vision works \cite{chen2015webly, fergus2010learning, xia2014well, divvala2014learning}. This applies for sound events as well and is, in fact, harder due to the complex and intricate ways we describe sounds \cite{kumar2018thesis}.  Often sound related terms are not mentioned in videos and hence text-based retrieval can lead to a much inferior collection of exemplars.

Nevertheless, a collection of exemplars for sound events obtained through automated methods will contain incorrect exemplars or label noise. This is a major challenge from the learning perspective. Learning from noisy labels has been a known problem \cite{frenay2014classification} and in recent years effectively training deep learning models from noisy labels has also started to get attention. However, it remains an open problem. Even here most of the work on learning from noisy labels has been in the domain of computer vision.

The next challenge is the presence of signal noise. Manual annotations keep in check the overall amount of signal noise in the data. Even if the source of data is the web (e.g AudioSet), manual labeling ensures that the sound is at least audible to most human subjects. However, for webly labeled data the signal noise is ``unchecked'' and the sound event, even if present, might be heavily masked by other sounds or noise. Signal noise in the webly labeled data is difficult to quantify and remains an open research topic for future works.

In this work, we develop an entire framework to deal with such webly labeled data. We begin by collecting a webly labeled dataset using a video search engine as the source. We then propose a deep learning based system for effectively learning from this webly labeled data. Our primary idea is that two neural networks can co-teach each other to robustly learn from webly labeled data. The two networks use two different views of the data due to which they have different learning capabilities. Since the labels are noisy, we argue that one cannot rely only on the loss with respect to the labels to train the networks. Instead, the agreement between the networks can be used to address this problem. Hence, we introduce a method to factor the agreement between the networks in the learning process. Our system also includes transfer learning to obtain robust feature representations. 

\section{Webly Labeled Learning of Sounds}
\label{sec:approach}

\subsection{Webly Labeled Training Data}

Obtaining training audio recordings is the first step in the learning process and is a considerable hard open problem on its own. The most popular approach in webly supervised systems in vision has been text query based retrieval from search engines \cite{fergus2010learning, xia2014well, chen2015webly}. Our approach is along similar lines where we use text queries to retrieve potential exemplars from a video search engine, YouTube.  


We must first select a ``vocabulary'' of sounds -- terms used to describe sounds. In this paper, we use a subset of 40 sound events from AudioSet, chosen based on several factors. These include preciseness in event names and definitions, the quality of metadata-based retrieval of videos from YouTube, the retention of sound hierarchies, and the number of exemplars in AudioSet (larger is better). 
\footnote{For more details visit github page mentioned in Section \ref{sec:expts}. }

\begin{figure}[t]
\centering
\includegraphics[width=0.95\columnwidth]{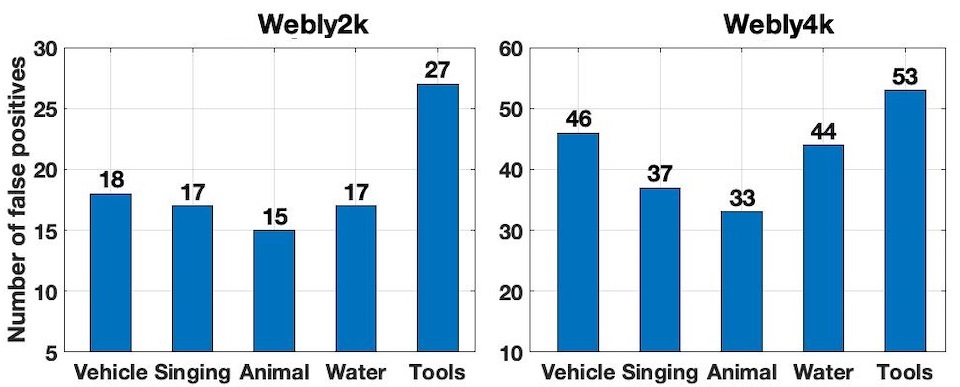}
\caption{False Positive for the 5 sound classes with the highest FP.}
\label{fig:fps}
\end{figure}
\subsubsection{Obtaining Webly Labeled Data}
Using only the sound name itself as a text query on YouTube leads to extremely noisy retrieval. \cite{kumar2017discovering} argued that humans often use the phrase ``sound of'' in texts before referring to a sound. Based on this intuition we augment the search query with the phrase ``sound of''. This leads to a dramatic improvement in the retrieval of relevant examples. For example, using \emph{``sound of dog''} instead of \emph{``dog''} improves the relevant results (sound event actually present in the recordings) by more than 60\% in the top 50 retrieved videos. Hence, we use the phrase ``sound of \textless sound-class\textgreater'' as the search query for retrieving example recordings of each class. 

We formed two datasets using the above strategy. The first one referred to as \textbf{\emph{Webly-2k}} uses top 50 retrieved videos for each class and has around 1,900 audio recordings. The second one, \textbf{\emph{Webly-4k}}, uses the top 100 retrieved videos for each class and contains around 3,800 recordings. Note that some recordings are retrieved for multiple classes, and hence, the datasets are multi-labeled, similar to AudioSet. Only recordings under 4 minutes duration are considered. 

\subsubsection{Analysis of the Dataset}
The average duration of the \emph{Webly-2k} set is around 111 seconds resulting in a total of around 60 hours of data.  Webly-4k is around 108 hours of audio with an average recording duration of 101 seconds. As mentioned before, label noise is expected in these datasets. To analyze this, we manually verified the positive exemplars of each class and estimated the number of false positives (FP) for each class. Clearly, the larger \emph{Webly-4k} contains far more noisy labels than \emph{Webly-2k}. 

Figure \ref{fig:fps} shows the FP counts for 5 classes with the highest false positives. Note that for these classes 30-50\% of the examples are wrongly labeled to contain the sound when it is actually not present. However, FP values can also be low for some classes, e.g., \emph{Piano} and \emph{Crowd}. Estimating false negatives requires one to manually check all of the recordings for all classes, which makes the task considerably difficult. Even the AudioSet dataset has not been assessed for false negatives (FN) and we also keep FN estimation out of scope of this paper. 

\begin{figure*}[t]
\centering
\includegraphics[width=0.90\linewidth]{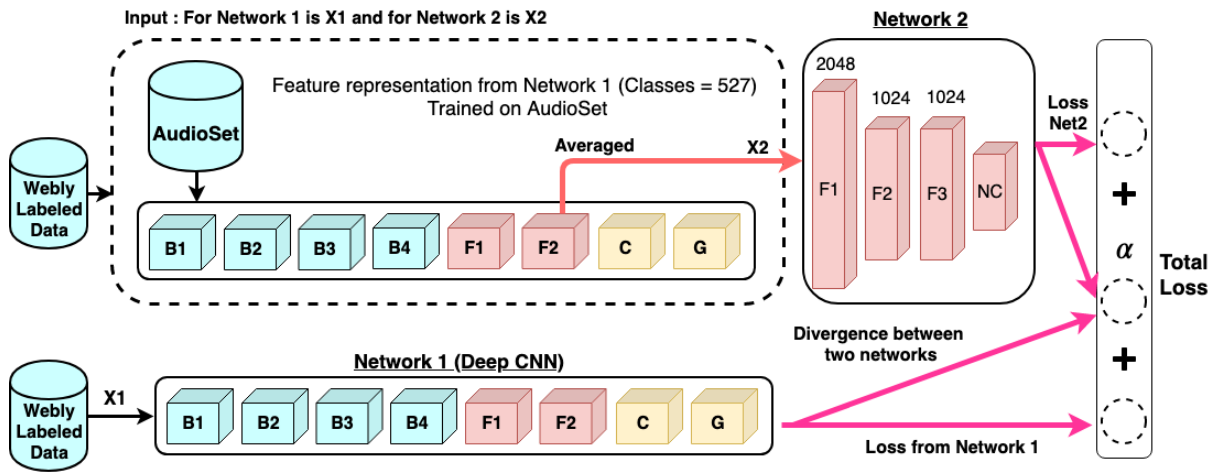}

\caption{WeblyNet System: Network 1 ($\mathcal{N}_1$) is a deep CNN with first view of data as input. Network 2 ($\mathcal{N}_2$) takes in the second view of data obtained through transfer learning. The networks are trained together to co-teach each other.}
\label{fig:audset}
\end{figure*}

\subsection{Proposed Approach: WeblyNet}
\label{ssec:approach}

The manual verification in the previous section was done only for analysis; the actual goal is to learn from noisily-labeled Webly-4k (or -2k) directly.
Robustly training neural networks with noisy labels remains a very hard problem \cite{tanaka2018joint}. Several methods have been proposed, especially in the visual domain \cite{goldberger2016training, chen2015webly, liang2016learning, reed2014training, tanaka2018joint, frenay2014classification}. These methods include bootstrapping, self-training, pseudo-labeling and curriculum learning, to mention a few. Another set of approaches try to estimate a noise transition matrix to estimate the distribution of noise in labels \cite{goldberger2016training}. However, estimating the noise transition matrix is not an easy problem as it is dependent on the representation and input features. Conventionally, ensemble learning has also been useful in handling noisy labels \cite{frenay2014classification}. 

In supervised learning, neural networks are trained on some divergence measure between the output produced by the network and the ground truth label. As the network is trained, the noise in the labels will lead to wrong updates in parameters which can affect the generalization capabilities of the network \cite{zhang2016understanding}. Some recent approaches have used the idea of having two networks working together to address this problem  \cite{malach2017decoupling,han2018co}.  \cite{malach2017decoupling} gives a ``when to update rule" where networks are updated when they disagree. In \cite{han2018co}, networks co-teach each other by sampling instances for training within a minibatch. 

Our approach is fundamentally based on the idea of training multiple networks together, where the agreement (or disagreement) between the networks are used for improved learning. The method incorporates ideas from co-training and multi-view learning. Multi-view learning methods (e.g., co-training \cite{blum1998combining}\cite{sun2013survey}) are primarily semi-supervised learning methods where learners are trained on different views of the data, and the goal is to maximize their agreement on the unlabeled data. Our proposed method exploits this central idea of the agreement between classifiers to address the challenges of webly labeled data. The intuition is as follows:  Two (or more) independent classifiers operating on noisily labeled data are likely to agree with the provided label when it is correct. When the given label is {\em incorrect}, the classifiers are unlikely to agree with it. They are, however, likely to agree with {\em one another} if both of them independently identify the correct label. Hence, the networks can inform each other on the errors they are making and help in filtering out those which are coming from noisy labels, thereby improving the overall robustness of the networks.

In contrast to prior work such as \cite{han2018co}, our proposed method explicitly ties in the co-teaching of the networks by having a disagreement measure in the loss term. Moreover, in our method, the two networks are operating on different views of the data and hence have different learning abilities. As a result, they will not fall in the degenerate situation where both networks essentially end up learning the same thing. This allows us to combine the classifiers' outputs during the prediction phase which further improves the performance. We refer to our overall system as WeblyNet. 

Furthermore, our method is easily extended to more than two networks. The central idea remains the same, a divergence measure captures disagreement measure between any two given pair of networks. Given $K$ networks in the system, $K(K-1)/2$ pairs of disagreements can be measured. These disagreement measures along with losses with respect to the available ground truths are then used to update the network parameters. Each divergence measure can be appropriately weighed by a scalar $\alpha$ to reflect the weight given to that particular pair of networks. Algorithm~\ref{alg:weblynet} outlines this procedure. 

\begin{algorithm}[tb]
   \caption{WeblyNet system}
       \textbf{Input}: Networks $\mathcal{N}_1$ to $\mathcal{N}_K$, Representation $R_1$ to $R_K$ of audio recordings for different networks and labels $Y$ of the recordings, learning rates $\eta_1$ to $\eta_K$, divergence weights $\alpha_1$ to $\alpha_{K(K-1)/2}$, number of epochs $n_{epochs}$ \\
    \textbf{Output}: Jointly trained networks
    \begin{algorithmic}[1]
    \For{$n$ = $1,2,....n_{epochs}$}
    \For{$k$ = $1,2,....K$}
        \State Compute loss, $\mathcal{L}_k(\mathcal{N}_k,Y)$ w.r.t label Y for network $\mathcal{N}_k$
    \EndFor
    \For{$k$ = $1,2,....K(K-1)/2$}
        \State Compute divergence $D(\mathcal{N}_i(R_i),\mathcal{N}_j(R_j))$ between each pair of networks
        \State Weigh each divergence by its corresponding hyperparameter $\alpha$
    \EndFor
        \State Combine all loss terms $\mathcal{L}()$ and divergence terms $D()$ 
        \State Update networks based on the combined loss. 
    \EndFor 
\end{algorithmic}
\label{alg:weblynet}
\end{algorithm}

In this work, we work with only two networks. Figure \ref{fig:audset} shows an overview of the proposed method. The two networks ``Network 1" and ``Network 2" take as input two different views of the data. The networks are trained jointly by combining their individual loss functions and a third divergence term which explicitly measures the agreement between the two networks. The individual losses provide supervision from given labels, and the mutual-agreement provides supervision when the labels are noisy. 

\subsubsection{Two Views of the Data} 
Our primary representation of audio recordings are embeddings provided by Google \cite{hershey2017cnn}. The embeddings are 128-dimensional quantized vectors for each 1 second of audio. Hence, an audio recording $\mathcal{R}$, in this first view,  is represented by a feature matrix $X_1 \in \mathcal{R}^{N X 128}$, where N depends on the duration of the audio. The temporal structure of the audio is maintained by stacking the embedding sequentially in $X_1$. The first network $\mathcal{N}_1$ is trained on these features. 

Several methods exist in the literature for generating multiple views from a single view \cite{sun2013survey}. In this work, we propose to use multiple non-linear transforms through a neural network to generate the second view ($X_2$) of the data. To this end, we use a network trained on first view, $X_1$, to obtain the second view of the data. This network is trained on another large scale sound events dataset (different from the webly labeled set we work with). One motivation is that, given the noisy nature of webly labeled data, a network trained on a large scale dataset such as Audioset can provide robust feature representations (as in transfer learning approaches \cite{kumar2017knowledge}). 

We first train a network ($\mathcal{N}_1$ with C = 527) on the AudioSet dataset and then use this trained model to obtain feature representations for our webly labeled data. More specifically, the F2 layer is used to obtain 1024-dimensional representations for the audio recordings by averaging the outputs across all 1-second segments. This representation learning through knowledge transfer, as shown empirically later, is significantly useful in webly labeled data where a higher level of signal noise and intra-class variation is expected. 

\subsubsection{Network Architectures: $\mathbf{\mathcal{N}_1}$ and $\mathbf{\mathcal{N}_2}$}
$\mathcal{N}_1$  is trained on the first ($X_1$) audio representations. It is a deep CNN. The layer blocks from B1 to B4 consists of two convolutional layers followed by a max-pooling layer. The number of filters in both convolutional layers of these blocks are, \{ B1:64, B2:128, B3:256, B4:256 \}. The convolutional filters are of size $3 \times 3$ in all cases, and the convolution operation is done with a stride of 1.  Padding of 1 is also applied to inputs of all convolutional layers. The max-pooling in these blocks are done using a window of size $1 \times 2$, moving by the same amount. Layer F1 and F2 are again convolutional layers with 1024 filters of size $1 \times 8$ and 1024 filters of size $1 \times 1$ respectively. All convolutional layers from B1 to F2 consists includes batch-normalization \cite{ioffe2015batch} and ReLU ($max(0,x)$) activations. The layer represented as $C$ is the segment level output layer. It consists of $C$ filters of size $1 \times 1$, where C is the number of classes in the dataset. This layer has a sigmoid activation function. The segment level outputs are pooled through a mapping function in the layer marked as $G$, to produce the recording level output. We use the average function to perform this mapping. 

The network architecture $\mathcal{N}_1$ achieves state-of-the-art results on AudioSet. In other words, the same architecture with  C = 527 (527 classes in AudioSet), achieves the  state-of-the-art result on AudioSet. Hence, we believe it is a good base network for our webly labeled learning. 

The network $\mathcal{N}_2$ (with $X_2$ as inputs) consists of 3 fully connected hidden layers with 2048, 1024 and 1024 neurons respectively. The output layer contains C number of neurons. A dropout of 0.4 is applied after first and second hidden layers. ReLU activation is used in all hidden layers and sigmoid in the output layer. 

\subsubsection{Training WeblyNet}
Given the multi-label nature of datasets, we first compute the loss with respect to each class. The output layer of both $\mathcal{N}_1$ and $\mathcal{N}_2$ gives posterior outputs for each class. We use the binary cross-entropy loss, defined with respect to $c^{th}$ class as, $l(y_c,p_c) = -y_c*log(p_c) - (1-y_c)*log(1-p_c)$. Here, $y_c$ and $p_c = \mathcal{N}(X)$ are the target and the network output for $c^{th}$ class, respectively. The overall loss function with respect to the target is the mean of losses over all classes, as shown in Eq \ref{eq:lossfn} 
\setlength{\belowdisplayskip}{0pt} \setlength{\belowdisplayshortskip}{0pt}
\setlength{\abovedisplayskip}{0pt} \setlength{\abovedisplayshortskip}{0pt}
\begin{equation}
\label{eq:lossfn}
\mathcal{L}(\mathcal{N}(X),y) = \frac{1}{C} \sum_{c=1}^C l(y_c,p_c)
\end{equation}

In the WeblyNet system, $\mathcal{N}_1$ and $\mathcal{N}_2$ co-teach each other through the following loss function
\begin{equation}
\label{eq:lossterm}
\begin{aligned}
\mathcal{L}(X_1,X_2,y) = \mathcal{L}(\mathcal{N}_1(X_1),y) + \mathcal{L}(\mathcal{N}_2(X_2),y) + \\ \alpha \cdot D(\mathcal{N}_1(X_1), \mathcal{N}_2(X_2))
\end{aligned}
\end{equation}

The first two terms in Eq\ref{eq:lossterm} are losses for the two networks with respect to the target. $D(\mathcal{N}_1(X_1), \mathcal{N}_2(X_2))$ is the divergence measure between the outputs of the two networks.  The divergence of “opinion” between the networks provides additional information beyond the losses with respect to target labels and helps reduce the impact of noisy labels on the learning process. The $\alpha$ term in Eq \ref{eq:lossterm} is a hyperparameter and controls the weight given to the divergence measure in the total loss. This can be set through a grid search and validation.

The divergence, $D(\mathcal{N}_1(X_1), \mathcal{N}_2(X_2))$, between the networks can be measured through a variety of functions. We found that the generalized KL-divergence worked best \cite{banerjee2005clustering}. Note that, the outputs from the two networks do not sum to 1. The generalized KL divergence is defined as $D_{KL}(\mathbf{x}||\mathbf{y}) = \sum_{i=1}^d x_i log(\frac{x_i}{y_i}) - \sum_{i=1}^d x_i + \sum_{i=1}^d y_i$. 
$D_{KL}(\mathbf{x}||\mathbf{y})$ is non-symmetric and is not a distance measure. We use $D(\mathbf{x},\mathbf{y}) = D_{KL}(\mathbf{x}||\mathbf{y}) + D_{KL}(\mathbf{y}||\mathbf{x})$ to measure the divergence between the outputs of the two networks. This measure is symmetric with respect to $\mathbf{x}$ and $\mathbf{y}$. If $\mathbf{o}^{\mathcal{N}_1}$ and $\mathbf{o}^{\mathcal{N}_2}$ are outputs from $\mathcal{N}_1$ and $\mathcal{N}_2$ respectively then $D(\mathcal{N}_1(X_1), \mathcal{N}_2(X_2))$ is 

\setlength{\belowdisplayskip}{0pt} \setlength{\belowdisplayshortskip}{0pt}
\setlength{\abovedisplayskip}{0pt} \setlength{\abovedisplayshortskip}{0pt}
\begin{equation}
\label{eq:gkld}
 D(\mathcal{N}_1(X_1), \mathcal{N}_2(X_2)) = \sum_{i=1}^C (\mathbf{o}^{\mathcal{N}_1}_i - \mathbf{o}^{\mathcal{N}_2}_i) log \frac{\mathbf{o}^{\mathcal{N}_1}_i}{\mathbf{o}^{\mathcal{N}_2}_i}
\end{equation}


During the inference stage, the prediction from WeblyNet is the average of the outputs from the networks $\mathcal{N}_1$ and $\mathcal{N}_2$.

\section{Experiments and Results}
\label{sec:expts}

WeblyNet is trained on the Webly-4k and Webly-2k training sets. Audio recordings are represented through $X_1$ and $X_2$ views (Sec. \ref{ssec:approach}). To the best of our knowledge, no other work has done an extensive study of webly supervised learning for sound events. Previous work, \cite{kumar2017audio}, is computationally not scalable to over 100 hours of data we use in this work. Moreover, it also relies on strongly labeled data in the learning process, which is not available in our case.  

All recordings from the \emph{Eval set} of AudioSet are used as the test set. This set contains around 4500 recordings corresponding to our set of 40 sound events. A subset of recordings from the \emph{Unbalanced set} of AudioSet is used for validation. Furthermore, to compare our webly supervised learning with a manually labeled set, we also create \emph{AudioSet-40} training set. \emph{AudioSet-40} is obtained from the \emph{balanced set} of the AudioSet by taking all recordings corresponding to the 40 sound events in our vocabulary with over 4,600 recordings.  

All experiments are done in PyTorch toolkit. Hyperparameters are tuned using the validation set. The network is trained through Adam optimization \cite{kingma2014adam}. Similar to other works \cite{gemmeke2017audio,kumar2017knowledge}, we use Average Precision (AP) as the performance metric for each class and then Mean Average Precision (MAP) of all classes as the metric for comparison. Please visit \textbf{\url{https://github.com/anuragkr90/webly-labeled-sounds}} for webly labeled data, codebase and additional analysis.

\begin{table}[t!]
  \centering
    \resizebox{1.0\columnwidth}{!}{
    \begin{tabular}{|c|c|c|c|}
    \hline
    \textbf{Method} & \textbf{MAP}  & \textbf{Method} & \textbf{MAP}  \\
    \hline
    WLAT  & \multirow{ 2}{*}{21.3} & ResNet-SPDA & \multirow{ 2}{*}{21.9}\\
    
    \cite{kumar2017knowledge} & &  \cite{zhang2016spda} & \\ 
    \hline
 
    ResNet-Attention&  \multirow{ 2}{*}{22.0} & ResNet-mean pooling  &  \multirow{ 2}{*}{21.8}\\
     \cite{xu2017attention} & & \cite{chou2018learning} & \\
    \hline
	M\&mmnet-MS & \multirow{ 2}{*}{22.6} & \multirow{ 2}{*}{\textbf{Ours} $\mathbf{\mathcal{N}_1}$} & \multirow{ 2}{*}{\textbf{22.9}}  \\ 
	 \cite{chou2018learning} & & & \\ 
    \hline
    \end{tabular}}
	\caption{MAP of $\mathcal{N}_{1}$ compared with state-of-the-art on whole AudioSet (527 Sound Events, Training: Balanced Set, Test: Eval Set)}  
  \label{tab:audres}%
\end{table}%

\paragraph{Full Audioset performance.} We begin by assessing the performance of $\mathcal{N}_1$ (with C=527) on AudioSet. The primary motivation behind this analysis is to show that the architecture of $\mathcal{N}_1$ is capable of obtaining state-of-the-art results on a standard well-known weakly labeled dataset. Table \ref{tab:audres} shows comparison with state-of-the-art. Our $\mathcal{N}_1$ is able to achieve state-of-art-performance on AudioSet. \cite{chou2018learning} reports a slightly better MAP of 23.2 using an ensemble of M\&mmnet-MS. Ensemble can improve our performance as well. The performance of $\mathcal{N}_1$ on AudioSet shows that it can serve as a good base architecture for our WeblyNet system. Hence, it also serves the baseline method for comparison. 

\begin{table}[t]
   \centering
\resizebox{0.49\columnwidth}{!}{
\begin{tabular}{|c|c|}
	\hline  
Methods & MAP \\
\hline
$\mathcal{N}_1$-Self (Baseline) (4k) & 38.7 \\
$\mathcal{N}_2$-Self (4k) & 41.4  \\

\textbf{WeblyNet} (4k) & \textbf{45.3}  \\
	\hline
\end{tabular}
}   
\resizebox{0.49\columnwidth}{!}{
\begin{tabular}{|c|c|}
	\hline  
Methods & MAP \\
\hline
$\mathcal{N}_1$-Self (Baseline) (2k) & 38.0 \\
$\mathcal{N}_2$-Self (2k) & 41.2  \\
\textbf{WeblyNet} (2k) & \textbf{44.0} \\
	\hline
\end{tabular}
}

\resizebox{0.99\columnwidth}{!}{   
\begin{tabular}{|c|c|c|c|}
	\hline  
Methods & MAP & Methods & MAP\\
\hline
$\mathcal{N}_1$-Self (Baseline) & 38.7 & $\mathcal{N}_1$ (Co-trained) & 43.6 \\
$\mathcal{N}_2$-Self & 41.4 & $\mathcal{N}_2$ (Co-trained) & 43.5 \\ 

$\mathcal{N}_1$-Self and $\mathcal{N}_2$-Self (Averaged) & 43.5 & WeblyNet & \textbf{45.3}   \\

$\mathcal{N}_2$ \cite{reed2014training} & 42.6 & &\\

\hline
\end{tabular}
}

\caption{Upper Tables: Comparison of systems on Webly-4k (L) and Webly-2k (R). Lower: $\mathcal{N}_1$ and $\mathcal{N}_2$ co-teach each other in WeblyNet leading to improvement in their individual performances over training them separately. Results shown on Webly-4k. See Sec. \ref{ssec:evalwebly}}    
\label{tab:weblymain}
\end{table}

\subsection{Evaluation of Webly Supervised Learning}
\label{ssec:evalwebly}

\begin{figure*}[h!]
    \centering
    \includegraphics[width=1.0\textwidth,height=0.405\textwidth]{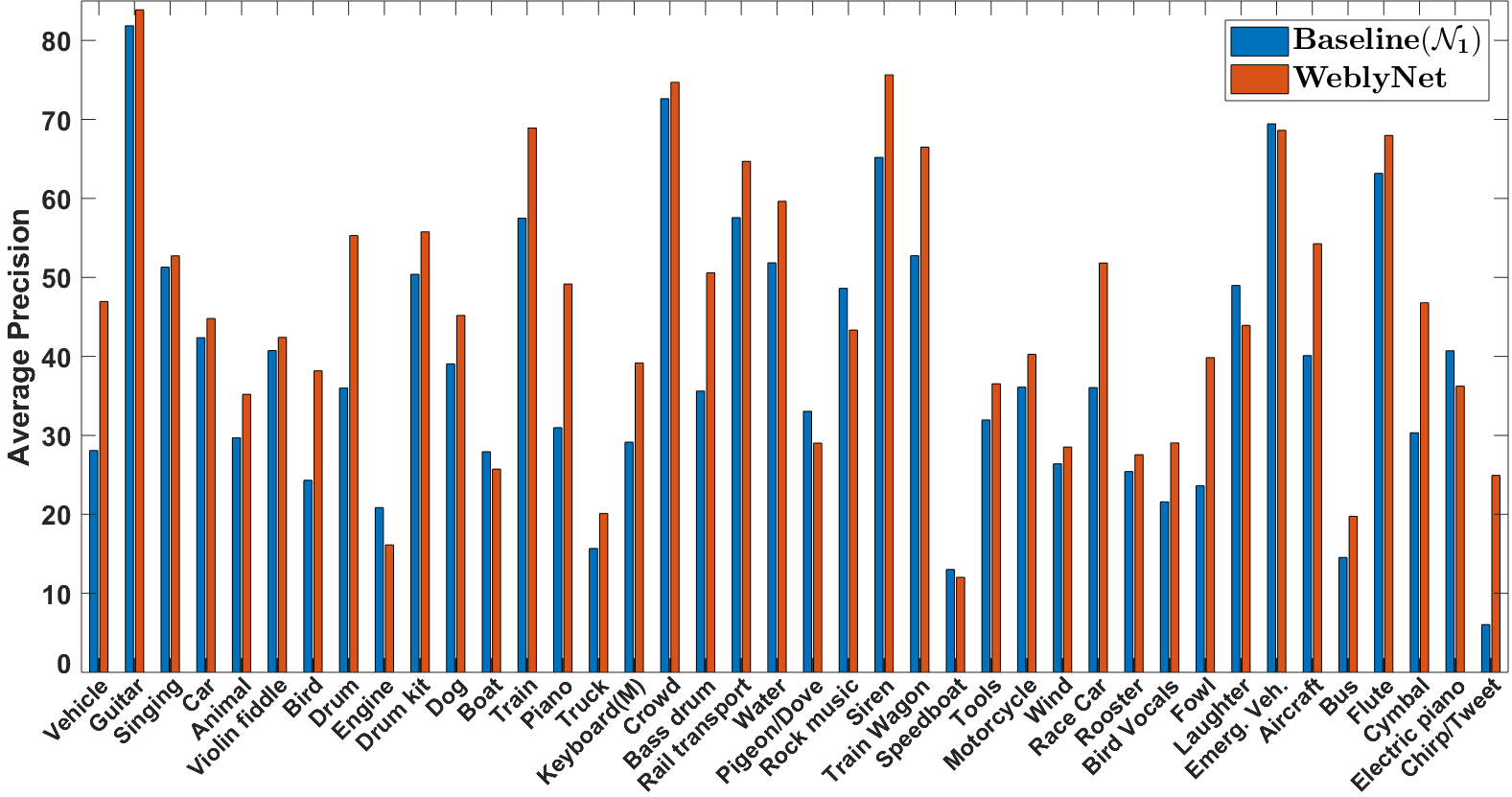}
    \caption{AP for sound events on Webly-4k training. Comparison of baseline ($\mathcal{N}_1$-Self) and WeblyNet System}
    \label{fig:comparison}
\end{figure*}

We first train $\mathcal{N}_1$ alone on the webly labeled dataset, and this performance ($\mathcal{N}_1$-Self) is taken as the baseline. We also train $\mathcal{N}_2$ alone on $X_2$ features ($\mathcal{N}_2$-Self) to assess the significance of the second view obtained through knowledge transfer in the webly supervised setting. To compare our method with a noisy label learning method, we apply the well-known approach described in \cite{reed2014training} for training $\mathcal{N}_2$. 

The upper two tables in Table-\ref{tab:weblymain} shows results for different systems on Webly-4k and Webly-2k training sets. We observe that WeblyNet leads to an absolute improvement of \textbf{6.6\%} (\textbf{17\%} relative) over the baseline method on the Webly-4k training set. Moreover, $X_2$ representations from pre-trained Audioset leads to considerable improvement over the baseline performance; around 7\% and 8.5\% relative improvements on the Webly-4k and Webly-2k training sets respectively. 

The lower table in Table-\ref{tab:weblymain} shows how our proposed system in which both networks co-teach each other leads to an improvement in the performance of the individual networks. First, we see that a simple combination of the two networks ($\mathcal{N}_1$-Self and $\mathcal{N}_2$-Self (Averaged) also leads to improved results. Once the WeblyNet system has been trained, we consider the output from individual $\mathcal{N}_1$ (or $\mathcal{N}_2$) as the output of the system. These are referred to as $\mathcal{N}_1$ (Co-trained) and $\mathcal{N}_2$ (Co-trained) respectively in Table \ref{tab:weblymain}. We can observe that the performances of both networks are improved by a considerable amount, over 12.7\% for $\mathcal{N}_1$ (38.7 to 43.6) and over 5\% for $\mathcal{N}_2$ (41.4 to 43.5). The overall WeblyNet system leads to 45.3 MAP. 

Moreover, the noisy label learning approach from \cite{reed2014training} leads to an improvement in $\mathcal{N}_2$ (to 42.6), 1\% absolute less than the improvement in $\mathcal{N}_2$ obtained with our co-training approach (43.6). This shows that ``learning together with agreement" is a more effective compared to bootstrapping in \cite{reed2014training}. Moreover, our overall system is 2.7\% absolute (6.4\% relative) better than that obtained through \cite{reed2014training}.

\begin{table}[t!]
  \centering
    \resizebox{1.0\columnwidth}{!}{
    \begin{tabular}{|l|l|l|l|l|l|}
    \hline
    Method & MAP  & Method & MAP & Method & MAP \\
    \hline
    AudioSet-40  & \multirow{ 2}{*}{54.3} & Webly-4k & \multirow{ 2}{*}{38.7} & Webly-4k & \multirow{ 2}{*}{\textbf{45.3}}\\
    $\mathcal{N}_1$-Self (Clean data) & & $\mathcal{N}_1$-Self  & & WeblyNet &  \\
    \hline
    \end{tabular}%
    }
	\caption{Webly labeled training vs. manual labeling (AudioSet-40)}  
  \label{tab:manual}%
\end{table}%

\begin{table}[t!]
  \centering
    \resizebox{0.85\columnwidth}{!}{
    \begin{tabular}{|l|l|l|l|l|}
    \hline
    Sounds & \multicolumn{2}{c|}{Webly-4k}  & \multicolumn{2}{c|}{Webly-2k} \\
    \cline{2-5}
       & $\mathcal{N}_1$-Self & WeblyNet & $\mathcal{N}_1$-Self & WeblyNet\\
      \hline
	Vehicle & 28.1 & 46.9 & 34.8 & 36.4 \\
	Singing & 51.3 & 52.7 & 47.8 & 50.5 \\
	Animal & 29.7 & 35.2 & 29.0 & 31.1 \\
	Water & 51.8 & 59.6 & 50.4 & 51.9 \\
	Tools & 31.9 & 36.5 & 40.7 & 40.6 \\
	\hline
	\textbf{Avg.} & \textbf{38.6} & \textbf{46.2} & \textbf{40.5} & \textbf{42.1} \\
    \hline
    \end{tabular}%
    }
	\caption{APs for 5 classes with high label noise in webly sets.}  
  \label{tab:classres}%
\end{table}%

\paragraph{Comparison with manual labeling.} Table \ref{tab:manual} shows comparison of $\mathcal{N}_1$ trained on AudioSet-40 (which is manually labeled) with systems trained on Webly-4k set.  Note, the test set is the same for all cases, only the training set is changing. A considerable difference of 15.6\%, exists between $\mathcal{N}_1$ trained on AudioSet-40 and that trained on Webly-4k. WeblyNet improves the webly supervised learning, by reducing this gap to 9.0\%. Such differences in performances between human-supervised data and non-human supervised webly data has been discussed in computer vision as well. 
Often, human supervision is hard to beat even by using even orders of magnitude more data \cite{chen2015webly}. 


\paragraph{Class specific results.}  Figure \ref{fig:comparison} shows the comparison of class-specific result comparison between baseline ($\mathcal{N}_1$-Self) and WeblyNet. WeblyNet improves over the baseline $\mathcal{N}_1$-Self for most of the classes (32 out of 40). Further, we analyze the performance of five classes with high label noise (Fig. \ref{fig:fps}). Table  \ref{tab:classres} shows the AP for these sounds. For all of these events, the improvements are considerable, e.g., around 67\% and 19\% relative improvements for Vehicle and Animal sounds, respectively. Interestingly, $\mathcal{N}_1$'s overall performance goes down for these 5 events as we increase the size of the dataset, from Webly-2k to Webly-4k. A considerable drop in performances are seen for Vehicle and Tools sounds whereas only small improvements Animal and Water sounds are seen. Deep learning methods are expected to improve as we increase the amount of training dataset. However, it is clear that the larger Webly-4k contains too many noisy labels which adversely affects the performances in certain cases. The proposed WeblyNet system is able to address this problem. On an average WeblyNet gives 7.6\% absolute (20\% relative) over the baseline method when trained on Webly-4k set. 

\paragraph{Effect of divergence measure.} To ensure that the divergence is playing a role in improving the system, we ran a sanity check experiment with $\alpha=0$  in Eq \ref{eq:lossterm}. The WeblyNet system, in this case, produces a MAP of 43.5, same as the simple combination of the individually trained networks. This is expected as the networks are not tied together anymore and one network will have no impact on the learning of the other.


\section{Conclusions}
\label{sec:conc}

Human supervision comes at a considerable cost, and hence we need to build methods which rely on human supervision to the least possible extent. In this paper, we presented webly supervised learning of sound events as the solution. We presented a method for mining web data and then a robust deep learning method to learn from webly labeled data. We showed that our proposed method in which networks co-teach each other leads to a considerable improvement in performance while learning from challenging webly labeled data. The method is extendable to more than two networks and in the future, we aim to explore this for efficient training of deep networks. Furthermore, we also need better methods to mine the web for sound events. This can involve clever natural processing techniques to associate metadata with different sound events and then assigning labels based on that. We keep this as part of our future works. 

\bibliographystyle{IEEEbib}
\footnotesize{\bibliography{ijcai19}}

\begin{thebibliography}{10}

\bibitem{kumar2018thesis}
Anurag Kumar,
\newblock {\em Acoustic Intelligence in Machines},
\newblock Ph.D. thesis, Carnegie Mellon University, 2018.

\bibitem{kumar2016audio}
Anurag Kumar and Bhiksha Raj,
\newblock ``Audio event detection using weakly labeled data,''
\newblock in {\em 24th ACM International Conference on Multimedia}. ACM
  Multimedia, 2016.

\bibitem{kumar2017knowledge}
Anurag Kumar, M.~Khadkevich, and C.~Fugen,
\newblock ``Knowledge transfer from weakly labeled audio using convolutional
  neural network for sound events and scenes,''
\newblock in {\em Acoustics, Speech and Signal Processing (ICASSP), 2018 IEEE
  International Conference on}. IEEE, 2018.

\bibitem{chou2018learning}
Szu-Yu Chou, Jyh-Shing~Roger Jang, and Yi-Hsuan Yang,
\newblock ``Learning to recognize transient sound events using attentional
  supervision.,''
\newblock in {\em IJCAI}, 2018, pp. 3336--3342.

\bibitem{mcfee2018adaptive}
Brian McFee, Justin Salamon, and Juan~Pablo Bello,
\newblock ``Adaptive pooling operators for weakly labeled sound event
  detection,''
\newblock {\em arXiv preprint arXiv:1804.10070}, 2018.

\bibitem{lu2018temporal}
Xugang Lu, Peng Shen, Sheng Li, Yu~Tsao, and Hisashi Kawai,
\newblock ``Temporal attentive pooling for acoustic event detection,''
\newblock {\em Proc. Interspeech 2018}, pp. 1354--1357, 2018.

\bibitem{xu2017attention}
Yong Xu, Qiuqiang Kong, Qiang Huang, Wenwu Wang, and Mark~D Plumbley,
\newblock ``Attention and localization based on a deep convolutional recurrent
  model for weakly supervised audio tagging,''
\newblock {\em arXiv preprint arXiv:1703.06052}, 2017.

\bibitem{gemmeke2017audio}
Jort~F Gemmeke, Daniel~PW Ellis, Dylan Freedman, Aren Jansen, Wade Lawrence,
  R~Channing Moore, Manoj Plakal, and Marvin Ritter,
\newblock ``Audio set: An ontology and human-labeled dataset for audio
  events,''
\newblock in {\em 2017 IEEE International Conference on Acoustics, Speech and
  Signal Processing (ICASSP)}. IEEE, 2017, pp. 776--780.

\bibitem{fonseca2017freesound}
Eduardo Fonseca, Jordi Pons~Puig, Xavier Favory, Frederic Font~Corbera, Dmitry
  Bogdanov, Andres Ferraro, Sergio Oramas, Alastair Porter, and Xavier Serra,
\newblock ``Freesound datasets: a platform for the creation of open audio
  datasets,''
\newblock in {\em Hu X, Cunningham SJ, Turnbull D, Duan Z, editors. Proceedings
  of the 18th ISMIR Conference; 2017 oct 23-27; Suzhou, China.[Canada]:
  International Society for Music Information Retrieval; 2017. p. 486-93.},
  2017.

\bibitem{chen2015webly}
Xinlei Chen and Abhinav Gupta,
\newblock ``Webly supervised learning of convolutional networks,''
\newblock in {\em Proceedings of the IEEE International Conference on Computer
  Vision}, 2015.

\bibitem{liang2016learning}
Junwei Liang, Lu~Jiang, Deyu Meng, and Alexander Hauptmann,
\newblock ``Learning to detect concepts from webly-labeled video data,''
\newblock IJCAI, 2016.

\bibitem{divvala2014learning}
Santosh~K Divvala, Ali Farhadi, and Carlos Guestrin,
\newblock ``Learning everything about anything: Webly-supervised visual concept
  learning,''
\newblock in {\em Proceedings of the IEEE Conference on Computer Vision and
  Pattern Recognition}, 2014, pp. 3270--3277.

\bibitem{kumar2017audio}
Anurag Kumar and Bhiksha Raj,
\newblock ``Audio event and scene recognition: A unified approach using
  strongly and weakly labeled data,''
\newblock in {\em Neural Networks (IJCNN), 2017 International Joint Conference
  on}. IEEE, 2017, pp. 3475--3482.

\bibitem{fergus2010learning}
Rob Fergus, Li~Fei-Fei, Pietro Perona, and Andrew Zisserman,
\newblock ``Learning object categories from internet image searches,''
\newblock {\em Proceedings of the IEEE}, vol. 98, no. 8, pp. 1453--1466, 2010.

\bibitem{xia2014well}
Yan Xia, Xudong Cao, Fang Wen, and Jian Sun,
\newblock ``Well begun is half done: Generating high-quality seeds for
  automatic image dataset construction from web,''
\newblock in {\em European Conference on Computer Vision}. Springer, 2014, pp.
  387--400.

\bibitem{frenay2014classification}
Beno{\^\i}t Fr{\'e}nay and Michel Verleysen,
\newblock ``Classification in the presence of label noise: a survey,''
\newblock {\em IEEE transactions on neural networks and learning systems},
  2014.

\bibitem{kumar2017discovering}
Anurag Kumar, Bhiksha Raj, and Ndapandula Nakashole,
\newblock ``Discovering sound concepts and acoustic relations in text,''
\newblock {\em submitted IEEE International Conference on Acoustics, Speech and
  Signal Processing (ICASSP)}, 2017.

\bibitem{tanaka2018joint}
Daiki Tanaka, Daiki Ikami, Toshihiko Yamasaki, and Kiyoharu Aizawa,
\newblock ``Joint optimization framework for learning with noisy labels,''
\newblock {\em arXiv preprint arXiv:1803.11364}, 2018.

\bibitem{goldberger2016training}
Jacob Goldberger and Ehud Ben-Reuven,
\newblock ``Training deep neural-networks using a noise adaptation layer,''
\newblock 2016.

\bibitem{reed2014training}
Scott Reed, Honglak Lee, Dragomir Anguelov, Christian Szegedy, Dumitru Erhan,
  and Andrew Rabinovich,
\newblock ``Training deep neural networks on noisy labels with bootstrapping,''
\newblock {\em arXiv preprint arXiv:1412.6596}, 2014.

\bibitem{zhang2016understanding}
Chiyuan Zhang, Samy Bengio, Moritz Hardt, Benjamin Recht, and Oriol Vinyals,
\newblock ``Understanding deep learning requires rethinking generalization,''
\newblock {\em arXiv preprint arXiv:1611.03530}, 2016.

\bibitem{malach2017decoupling}
Eran Malach and Shai Shalev-Shwartz,
\newblock ``Decoupling" when to update" from" how to update",''
\newblock in {\em Advances in Neural Information Processing Systems}, 2017, pp.
  960--970.

\bibitem{han2018co}
Bo~Han, Quanming Yao, Xingrui Yu, Gang Niu, Miao Xu, Weihua Hu, Ivor Tsang, and
  Masashi Sugiyama,
\newblock ``Co-teaching: robust training deep neural networks with extremely
  noisy labels,''
\newblock 2018.

\bibitem{blum1998combining}
Avrim Blum and Tom Mitchell,
\newblock ``Combining labeled and unlabeled data with co-training,''
\newblock in {\em Proceedings of the eleventh annual conference on
  Computational learning theory}. ACM, 1998, pp. 92--100.

\bibitem{sun2013survey}
Shiliang Sun,
\newblock ``A survey of multi-view machine learning,''
\newblock {\em Neural Computing and Applications}, vol. 23, no. 7-8, pp.
  2031--2038, 2013.

\bibitem{hershey2017cnn}
Shawn Hershey, Sourish Chaudhuri, Daniel~PW Ellis, Jort~F Gemmeke, Aren Jansen,
  R~Channing Moore, Manoj Plakal, Devin Platt, Rif~A Saurous, Bryan Seybold,
  et~al.,
\newblock ``Cnn architectures for large-scale audio classification,''
\newblock in {\em 2017 IEEE International Conference on Acoustics, Speech and
  Signal Processing (ICASSP)}. IEEE, 2017, pp. 131--135.

\bibitem{ioffe2015batch}
Sergey Ioffe and Christian Szegedy,
\newblock ``Batch normalization: Accelerating deep network training by reducing
  internal covariate shift,''
\newblock {\em arXiv preprint arXiv:1502.03167}, 2015.

\bibitem{banerjee2005clustering}
Arindam Banerjee, Srujana Merugu, Inderjit~S Dhillon, and Joydeep Ghosh,
\newblock ``Clustering with bregman divergences,''
\newblock {\em Journal of machine learning research}, vol. 6, no. Oct, 2005.

\bibitem{kingma2014adam}
Diederik~P Kingma and Jimmy Ba,
\newblock ``Adam: A method for stochastic optimization,''
\newblock {\em arXiv preprint arXiv:1412.6980}, 2014.

\bibitem{zhang2016spda}
Han Zhang, Tao Xu, Mohamed Elhoseiny, Xiaolei Huang, Shaoting Zhang, Ahmed
  Elgammal, and Dimitris Metaxas,
\newblock ``Spda-cnn: Unifying semantic part detection and abstraction for
  fine-grained recognition,''
\newblock in {\em Proceedings of the IEEE Conference on Computer Vision and
  Pattern Recognition}, 2016, pp. 1143--1152.

\end{thebibliography}
\end{document}